\documentclass[twocolumn,showpacs,amsfonts,amsmath,amssymb,superscriptaddress]{revtex4}
\usepackage{graphics}
\usepackage{dcolumn}
\usepackage{bm}

\begin{document}

\title{Realization of Fredkin Gate by Three Transition Pulses \\in NMR Quantum Information Processor}

\affiliation{Department of Modern Physics, University of Science
and Technology of China,
  Hefei, People's Republic of China, 230027\\}
\affiliation{Laboratory of Structure Biology, University of
Science and Technology of China,
  Hefei, People's Republic of China, 230027}

\author{\surname{XUE} Fei}
\email{Feixue@mail.ustc.edu.cn}
\author{\surname{DU} JiangFeng}
\author{\surname{SHI} MingJun}
\author{\surname{ZHOU} XianYi}
\author{\surname{HAN} RongDian}
\affiliation{Department of Modern Physics, University of Science
and Technology of China,
  Hefei, People's Republic of China, 230027\\}
\author{\surname{WU} JiHui}
\affiliation{Laboratory of Structure Biology, University of
Science and Technology of China,
  Hefei, People's Republic of China, 230027}

\date{Apr 19, 2002}

\begin{abstract}
The three-qubit conditional swap gate(Fredkin gate) is a universal
gate that can be used to create any logic circuit and has many
direct usages. In this paper, we experimentally realized Fredkin
gate with only three transition pulses in solution of alanine. It
appears that no experimental realization of Fredkin gate with
fewer pulses has yet been reported up to now. In addition, the
simple structure of our scheme makes it easy to be implemented in
experiments.

\end{abstract}


\maketitle

Quantum information processing offers great advantages over
classical information processing, both for efficient
algorithms\cite{Shor,Grover,Daniel} and for secure communication
\cite{Bennett}. It can be accomplished by universal gates, such as
Fredkin gate, Toffoli gate and the controlled-not(CNOT) gate. Of
them Fredkin gate is of interest because it is not only a
universal gate for classical reversible
computation\cite{Fredkin}, but also has direct applications in
error correcting\cite{Ekert}, polarization transfer in
NMR\cite{DE.Chang} and some quantum algorithm\cite{Buhrman}.
Various schemes have been proposed to implement Fredkin gate. In
principle any multi-qubit gate can be build up from combination
of the CNOT gates and the single-qubit
gates\cite{Barenco,Sleator}. Chau and Wilczek have given a
construction of Fredkin gate with six specific gates\cite{Chau}.
Then Smolin and DiVicenzo show that five two-qubit gates are
sufficient to implement it\cite{Smolin}, and no scheme with
smaller number of two-qubit gates was reported yet. Of the
growing number of methods\cite{Cory1,Cirac,Kane} proposed to
realize Quantum Information Processor(quantum computer),
NMR(Nuclear Magnetic Resonance) Quantum Information
Processor(QIP)\cite{Cory2} has given the most successful
result\cite{Lieven1,Lieven2,Knill,Collins}. In NMR QIP quantum
gates are realized by pulse sequences. For example the CNOT gate
was realized by a pulse sequence composed of five
pulses\cite{Gershenfeld}. Then the pulse sequence of Fredkin gate
using two-qubit gates would include over twenty pulses in NMR
QIP. The more pulses we used in the sequence, the more complexity
and the more errors induced in experiment. Efforts are made to
reduce the number of pulses in the sequences and to simplify the
structure of the pulse sequences realizing the quantum
gates.\cite{Du1,Du2,Price}

\begin{figure}[bp]
\centering
\includegraphics{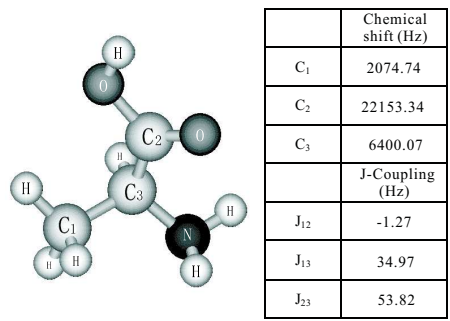}
\caption{\label{alanine&table}The structure of alanine and the
table of the chemical shifts and J-coupling constants, the
chemical shifts are given with respect to reference frequency
125.76MHz(carbons) on Bruker Avance DMX500 spectrometer. The
three weakly coupled spin$\frac{1}{2}$ carbon-13 nuclei served as
three qubits, which are labeled by $C_1, C_2, C_3$.}
\end{figure}

In this paper, we report a practical experimental realization of
Fredkin gate, which is accomplished by only three transition
pulses in NMR QIP. The molecule that we use is alanine(FIG.
\ref{alanine&table}). The effect of undesired coupling on the H
nuclei which is also spin$\frac{1}{2}$ is removed by common
selective decoupling technique in NMR. Our experimental results
show that this scheme is convenient and easy to be realized. And
it appears that no experimental realization of Fredkin gate with
fewer pulses has yet been reported up to now.

\begin{figure}[!]
\centering
\includegraphics{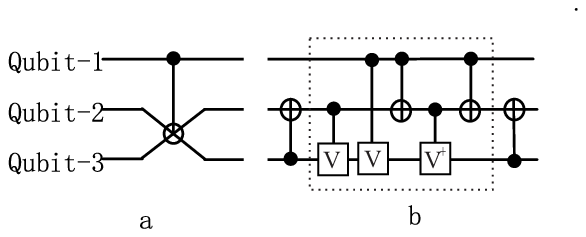}
\caption{\label{fredkingates} a. Quantum circuit symbol for
Fredkin gate.  b. Smolin and DiVincenzo's construction of Fredkin
gate with seven two-qubit gates, the gates in the dot-lined frame
is the Toffoli gate.}
\end{figure}

Fredkin gate is also called controlled-swap gate(FIG.
\ref{fredkingates}a), that is, when the controlling qubit(the
first qubit) is in the state $\vert1\rangle$, the two controlled
qubits(the second and third qubits) exchange their states after
the action of Fredkin gate. Otherwise, if the controlling qubit
is in state $\vert0\rangle$, the two controlled qubits remain
their original states. Smolin and DiVincenzo have suggested a
scheme to implement Fredkin gate(FIG.
\ref{fredkingates}b)\cite{Smolin}. Their scheme depends greatly on
implementation of two-qubit quantum gates and the combination of
them. But in NMR QIP implementing two-bit gate in a multi-spin
system is more complex than implementing it in two-spin system.
In other words, in NMR QIP, for multi-spin system combining
two-qubit gates is not always an efficient way to implement
bigger gates. The reason is that considering two-qubit gates as
the basic elements is mainly for mathematics convenience,and can
help us calculate and understand the general questions in quantum
information. However, in NMR QIP considering two-qubit gates as
the basic gates is not a reasonable choice, a better choice would
be fewer pulses and the pulse sequences which have simple
structure and are easier to be realized.

\begin{figure}[!]
\centering
\includegraphics{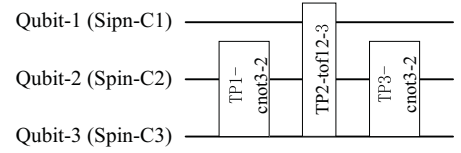}
\caption{\label{pulsefredkin} The pulse sequence for Fredkin
gate, which only includes three transition pulses.}
\end{figure}

So we reconsider Smolin and DiVincenzo's scheme to implement
Fredkin gate that is composed of two CNOT gates and one Toffoli
gate(FIG. \ref{fredkingates}b). We do not keep on decomposing the
Toffoli gate into two-qubit gates. However, inspired by the method
Du and his collaborators suggested\cite{Du2}, we notice that
Fredkin gate may be realized with only three transition pulses.
Transition pulses are deliberately designed to perturb transverse
magnetic field to sway only a small area of spectra. The pulse
sequence is shown in FIG. \ref{pulsefredkin}, and the parameters
of the transition pulses are shown in TABLE.\ref{parameter}.
TP$i$-cont3-2($i=1,3$) is a $\pi$ transition pulse to implement
the CNOT gate(qubit-3 as controlling qubit and qubit-2 as
controlled qubit).
\begin{eqnarray}
\label{Tcnot32}
U_{TP1}=exp[i\pi\frac{1}{4}\sigma_y^2(1-\sigma_z^3) \\
=
\begin{pmatrix}
  1 & 0 & 0 & 0 & 0 & 0 & 0 & 0 \\
  0 & 0 & 0 & 1 & 0 & 0 & 0 & 0 \\
  0 & 0 & 1 & 0 & 0 & 0 & 0 & 0 \\
  0 & -1 & 0 & 0 & 0 & 0 & 0 & 0 \\
  0 & 0 & 0 & 0 & 1 & 0 & 0 & 0 \\
  0 & 0 & 0 & 0 & 0 & 0 & 0 & 1 \\
  0 & 0 & 0 & 0 & 0 & 0 & 1 & 0 \\
  0 & 0 & 0 & 0 & 0 & -1 & 0 & 0
\end{pmatrix} \\
U_{TP3}=exp[-i\pi\frac{1}{4}\sigma_y^2(1-\sigma_z^3)] \\
=
\begin{pmatrix}
  1 & 0 & 0 & 0 & 0 & 0 & 0 & 0 \\
  0 & 0 & 0 & -1 & 0 & 0 & 0 & 0 \\
  0 & 0 & 1 & 0 & 0 & 0 & 0 & 0 \\
  0 & 1 & 0 & 0 & 0 & 0 & 0 & 0 \\
  0 & 0 & 0 & 0 & 1 & 0 & 0 & 0 \\
  0 & 0 & 0 & 0 & 0 & 0 & 0 & -1 \\
  0 & 0 & 0 & 0 & 0 & 0 & 1 & 0 \\
  0 & 0 & 0 & 0 & 0 & 1 & 0 & 0
\end{pmatrix}
\end{eqnarray}
The phases of TP1 and TP3 are set to be inverse with each other.
It would help to greatly conceal undesired effects of the
chemical shifts evolution during the length of TP2-tof-12-3 and
the time needed to switch the on-resonance frequencies($O_1$).
TP2-tof-12-3 is a $\pi$ transition pulse to implement the Toffoli
gate(qubit-1,2 as controlling qubit, qubit-3 as controlled qubit),
\begin{eqnarray}
\label{Ttof123}
U_{TP2}=exp[-i\pi\frac{1}{8}(1-\sigma_z^1)(1-\sigma_z^2)\sigma_x^3]
\\
=
\begin{pmatrix}
  I_6 &  0 \\
  0 & -i\sigma_x
\end{pmatrix}
\end{eqnarray}
The three transition pulses are all 180-degree pulses. This is
also help to reduce undesired effects of the chemical shifts
during the length of transition pulses and J-couplings evolution
during the length of transition pulses. Combining the three
transition pulses we get
\begin{eqnarray}
U_{Fred}=U_{TP3}.U_{TP2}.U_{TP1}\nonumber \\
=\begin{pmatrix}
  I_5 & 0 & 0 \\
    0 & -i\sigma_x & 0 \\
    0 & 0 & 1 \\
\end{pmatrix}
\end{eqnarray}

\begin{table}[!]
\caption{\label{parameter} The Parameters of the Three Transition
Pulses. All pulses have the length of 66.56ms and are square
shape.}
\begin{ruledtabular}
\begin{tabular}{cccc}
Transition Pulses& $O_1$(Hz) &shift(Hz)&Phase(degree)\\
\hline
TP1-cont3-2&22153.34&-26.91&0\\
TP2-tof-12-3&6400.07&-44.90&90\\
TP3-cont3-2&22153.34&-26.91&180\\
\end{tabular}
\end{ruledtabular}
\end{table}

To confirm whether Fredkin gate is correctly realized, We act the
pulse sequence on the state
\begin{equation}
\rho_{in}=\frac{1}{2^3}I_{8\times8}+\epsilon(I^1_x+I^2_z+I^3_x)
{\label{math:in}}
\end{equation}
where $I^k_\alpha(k={1,2,3}, \alpha={x,y,z})$ is the matrix for
the $\alpha$-component of the angular momentum of the spin-$k$.
The relation between $I^k_\alpha$ and Pauli Matrix
$\sigma^k_\alpha$ is $I^k_\alpha=\sigma^k_\alpha/2$. This state
can be obtained from the equilibrium state
\begin{equation}
\rho_{eq}=\frac{1}{2^3}I_{8\times8}+\epsilon(I^1_z+I^2_z+I^3_z)
{\label{math:eq}}
\end{equation}
by applying two selective pulse $R^1_y(\frac{\pi}{2})$ and
$R^3_y(\frac{\pi}{2})$. Calculation shows that the final state
should be
\begin{eqnarray}
\rho_{out}=U_{TP3}.U_{TP2}.U_{TP1}.\rho_{in}.U_{TP1}^\dagger.U_{TP2}^\dagger.U_{TP3}^\dagger\nonumber \\
=\frac{1}{2^3}I_{8\times8}+\frac{\epsilon}{2}(I^1_x+I^3_x+I^2_z+I^3_z \nonumber \\
+2I^1_zI^3_x-2I^2_yI^3_z+2I^1_zI^2_z-2I^1_zI^3_z \nonumber\\
+4I^1_xI^2_zI^3_z+4I^1_zI^2_yI^3_z-4I^1_yI^2_xI^3_x-4I^1_yI^2_yI^3_y)
{\label{math:final}}
\end{eqnarray}

\begin{figure}[!]
\centering
\includegraphics{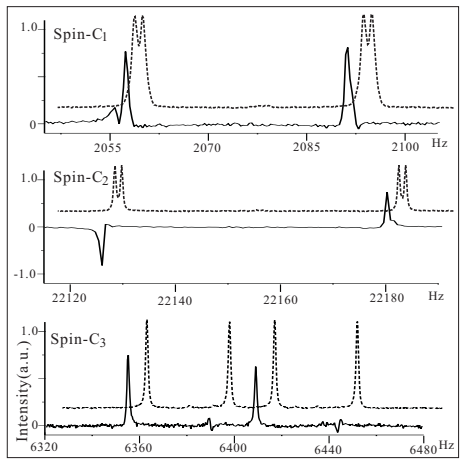}
\caption{\label{EXP_spec} The experiment spectra, the solid-lined
spectra are from state $\rho_{out}$ without any readout pulse(the
spectra of Spin-$C_2$ with 90 degree phase adjusting); the
dot-lined spectra are from the state $\rho_{eq}$ with readout
pulse $R^{1,2,3}_y(\frac{\pi}{2})$}
\end{figure}

The experimental spectra corresponding to the state of
$\rho_{out}$ without applying any readout pulse are shown in
FIG.\ref{EXP_spec}. There are two inner peaks in the spectra of
qubit-1(Spin-$C_1$), which is the expected spectra form of
$I^1_x+4I^1_xI^2_zI^3_z$; two left peaks(inverse phase) of
$J_{12}$ in the spectra qubit-2(Spin-$C_2$), which is the expected
spectra form of $-2I^2_yI^3_z+4I^1_zI^2_yI^3_z$ with 90 degree
phase adjusting(Notice that $J_{12}$ has a different sign with
other J.); two left peaks of $J_{13}$ in the spectra
qubit-3(Spin-$C_3$), which is the expected spectra form of
$I^3_x+2I^1_zI^3_x$. These spectra together show that we gain the
expected state $\rho_{out}$ after applying the three transition
pulses to the state $\rho_{in}$. We have also tried other states
as the input state and also gained the corresponding expected
output states. Experiment results implies that the scheme is
working for all 8 basic product basis states. Hence we conclude
that Fredkin gate is successfully realized by the three
transition pulses.

In conclusion, with a solution of alanine, we have experimentally
demonstrated that Fredkin gate can be realized with only three
transition pulses in NMR QIP. The pulse sequence that we use has
a excellent symmetry. Such symmetry makes the realization of
Fredkin gate easier to accomplish in experiment, and also makes
the realization be more accurate. We notice that there are small
bumps in the spectra of qubit-1(Spin-$C_1$) and the spectra
qubit-2(Spin-$C_2$). These distortions are really small if we
notice that the resolution of our Bruker Avance DMX500
spectrometer(0.5Hz) and $J_{12}$(1.27Hz) are comparable. However
small, they show that our realization of Fredkin gate is not
perfect. Reasons are: first, the magnetic field is not
homogeneous; second, $\pi$ transition pulses rotate not precisely
180 degree in experiment; third, although chemical shifts
evolution during the length of TP2-tof-12-3 and the time needed
to switch the on-resonance frequencies($O_1$) are greatly
concealed, there are still a little undesired effects of the
chemical shifts left during the time of pulse sequence. Our
realization of Fredkin gate depend on the realization of
transition pulses. For other molecule whose $J_{12}$ is bigger,
realizing TP$i$-cont3-2 will need to implement two single-line
selective pulses. With the technique presented by Steffen, Lieven
and Chuang in paper \cite{Steffen} it can be accomplished with
satisfying accuracy at the same time. So our scheme is also
practicable in other molecules, and in principle it is ready to
use in NMR QIP.

This work was supported by the National Nature Science Foundation
of China(Grants No. 10075041 and 10075044).

\end{document}